\documentclass{article}

\usepackage{arxiv}

\usepackage{graphicx}
\usepackage{subfigure} 
\usepackage[utf8]{inputenc} 
\usepackage[T1]{fontenc}    
\usepackage{hyperref}       
\usepackage{url}            
\usepackage{booktabs}       
\usepackage{amsfonts}       
\usepackage{nicefrac}       
\usepackage{microtype}      
\usepackage{lipsum}
\usepackage{graphicx}
\graphicspath{ {./images/} }
\usepackage{amsmath}

\title{Integrating Protein Sequence and Expression Level to Analyze Molecular Characterization of Breast Cancer Subtypes}

\author{
Hossein Sholehrasa\textsuperscript{1,2} and
Majid Jaberi-Douraki\textsuperscript{1,3}\thanks{Corresponding Author: jaberi@k-state.edu}\\
\textsuperscript{1}1DATA Consortium and FARAD Program, Kansas State University, Olathe, KS, USA\\
\textsuperscript{2}Department of Computer Science, Kansas State University, Manhattan, KS, USA\\
\textsuperscript{3}Department of Mathematics, Kansas State University, Manhattan, KS, USA\\
}

\begin{document}
\maketitle
\begin{abstract}
Breast cancer's complexity and variability pose significant challenges in understanding its progression and guiding effective treatment. This study aims to integrate protein sequence data with expression levels to improve the molecular characterization of breast cancer subtypes and predict clinical outcomes. Using ProtGPT2, a language model specifically designed for protein sequences, we generated embeddings that capture the functional and structural properties of proteins. These embeddings were integrated with protein expression levels to form enriched biological representations, which were analyzed using machine learning methods, such as ensemble K-means for clustering and XGBoost for classification. Our approach enabled the successful clustering of patients into biologically distinct groups and accurately predicted clinical outcomes such as survival and biomarker status, achieving high performance metrics, notably an F1 score of 0.88 for survival and 0.87 for biomarker status prediction. Feature importance analysis identified KMT2C, CLASP2, and MYO1B as key proteins involved in hormone signaling, cytoskeletal remodeling, and therapy resistance in hormone receptor-positive and triple-negative breast cancer, with potential influence on breast cancer subtype behavior and progression. Furthermore, protein-protein interaction networks and correlation analyses revealed functional interdependencies among proteins that may influence the behavior and progression of breast cancer subtypes. These findings suggest that integrating protein sequence and expression data provides valuable insights into tumor biology and has significant potential to enhance personalized treatment strategies in breast cancer care.

\end{abstract}


\section{Introduction}
Breast cancer is still a major global health issue, making up a significant number of cancer cases and deaths among women \cite{Biomarker_luminal}. Despite advancements in early detection and treatment, the complexity and diversity of breast cancer present challenges in understanding its progression, predicting outcomes, and creating effective treatment plans. The classification of breast cancer into various subtypes based on molecular and clinical features is a crucial aspect of guiding treatment decisions and improving patient outcomes. Mainly, the presence or absence of specific biological markers, such as Estrogen Receptor (ER), Progesterone Receptor (PR), and Human Epidermal Growth Factor Receptor 2 (HER2), serves as a foundation for categorizing breast cancer into luminal, HER2-enriched, or basal-like subtypes \cite{ planes2019pd, cheang2009ki67}. These markers are essential in determining how a tumor may respond to hormone therapy or other targeted treatments, making their accurate classification vital for personalized medicine \cite{mittal2023molecular}.

Recent advances in proteomics and genomics have significantly enhanced the molecular understanding of breast cancer. Proteomics, the large-scale study of proteins, adds a critical layer to cancer research since proteins are the functional entities in cells that carry out biological processes \cite{hanash2003disease}. In breast cancer, shifts in protein expression patterns can provide valuable insights into tumor behavior, disease progression, and potential therapeutic targets. For instance, ER, PR, or HER2 changes can indicate how aggressive a tumor might be and what treatment strategies may be most effective. Understanding these markers not only aids in diagnosis but also plays a crucial role in therapeutic decision-making \cite{mittal2023molecular, litvin2020prognosis}.

One of the key innovations in this work involves integrating protein expression data with the underlying protein sequences. Protein expression levels reflect the abundance of specific proteins within cells, shaped by processes such as translation and degradation, which are critical in regulating cellular functions. These dynamics play a significant role in determining cellular behavior, including in tumor contexts, where protein turnover and regulation influence functional outcomes and adaptations  \cite{munro2024cellular}. On the other hand, protein sequences, through their structural determinants, provide insights into functional domains, enabling a deeper understanding of their roles in biological processes. This sequence-structure-function relationship forms the foundation for exploring how specific protein functions emerge, offering a pathway to study processes such as cancer progression by integrating structural and functional annotations \cite{koehler2023sequence}. Combining these two layers of information, protein expression level, and sequence features, we can capture both quantitative and qualitative aspects of the proteins, leading to a more comprehensive understanding of their roles in breast cancer biology.

We used ProtGPT2 \cite{ferruz2022protgpt2}, an advanced large language model (LLM) based on the GPT-2 architecture, to facilitate the integration of protein sequence and expression data. ProtGPT2 is trained explicitly for protein sequences, enabling it to generate meaningful embeddings that represent proteins' functional and structural properties. These embeddings provide a high-dimensional representation that captures complex details of each protein sequence, allowing the model to leverage both local and global features of the sequences \cite{ferruz2022protgpt2}. Integrating transcriptomic data with protein expression levels in breast cancer enables the construction of a biologically comprehensive representation of tumor biology, facilitating downstream analyses such as clustering tumors into proteomic subgroups and associating these clusters with patient survival outcomes and molecular subtypes \cite{tang2018integrated}. Our approach, which incorporates machine learning methods like ensemble K-means for clustering and classification algorithms such as XGBoost and Random Forest, aims to uncover relationships between protein features and clinical characteristics, offering new insights into the grouping of breast cancer patients and potential predictive biomarkers. Furthermore, the importance of trust and explainability in using artificial intelligence (AI) for healthcare has been highlighted in recent studies, including \cite{10.1007/978-3-031-61966-3_37} that explore mechanisms to bridge the gap between advanced AI technologies and clinical practice, specifically in the breast cancer. In this context, the integration of protein sequence embeddings with expression levels may be considered a biologically meaningful framework for understanding protein function. Coupled with machine learning methods, enables the categorization of breast cancer patients and identification of potential predictive biomarkers.

This research aims to integrate protein sequence and expression data to enhance our understanding of tumor subtypes and predict clinical outcomes for breast cancer patients. The study focuses on understanding the interactions among vital proteins and clinical features, such as tumor size and patient age, to develop robust models for patient classification and survival prediction.

\section{Methodology}
\subsection{Data Source}
The dataset used in this study is derived from the proteomic data published by \cite{grabowski2017breastcancer, mertins2016proteogenomics}. The authors selected 105 breast tumor samples previously characterized by The Cancer Genome Atlas (TCGA) for proteomic analysis. Then, these samples represented various mRNA expressions of 50 genes (PAM50) intrinsic subtypes, including basal-like, luminal A, luminal B, and HER2-enriched tumors, along with normal breast tissue samples. The authors analyzed the samples using high-resolution accurate-mass tandem mass spectrometry (MS/MS), with extensive peptide fractionation and phosphopeptide enrichment. In the next step, the protein and phosphosite levels were quantified using the isobaric tag for relative and absolute quantification (iTRAQ) method. After the authors did rigorous quality control, 28 samples were excluded due to protein degradation, resulting in 77 tumor samples and 12,553 proteins, which were used in this study along with clinical data, including patient age and tumor size \cite{mertins2016proteogenomics}.

In our work, we utilized the dataset of 12,553 proteins from the TCGA proteomics and then extracted protein sequence data from the National Center for Biotechnology Information (NCBI) database \cite{ncbi}. The protein sequence data were retrieved using API calls based on the RefSeq IDs provided in the TCGA dataset. The extracted protein sequences were obtained in Fast-All (FASTA) format, ensuring compatibility with further bioinformatic and proteomic analyses. This additional data collection allowed for integrating protein sequence information with the proteomic expression levels, enriching the dataset used in our study.

\subsection{Preprocessing}
The clinical data used in this analysis includes various attributes such as tumor size, lymph node involvement (Node), metastasis status, and other relevant features. Label encoding was applied to prepare the categorical variables for analysis. Label encoding transforms these categorical variables into numeric values, making them suitable for machine learning algorithms that require numerical input. For patient age, a min-max scaling was employed for normalization. This technique scales the data by rescaling the values between 0 and 1. The advantage of using the min-max scaling is that it preserves the relationships between data points while ensuring that no variable dominates others due to differing scales.

Regarding protein expression levels, proteins with more than 30\% missing values in patient data were removed from the dataset. This filtering step was necessary to improve data quality and avoid bias from incomplete data. As a result, the number of proteins was reduced from 12,553 to 10,625. The remaining proteins with missing values were imputed using the median protein expression levels across patients. The median was chosen as the imputation method because it is robust to outliers and provides a more stable central tendency than the mean, which can skew extreme values. By filling the missing data with the median, the overall structure of the dataset remains more consistent.

\subsection{Integrating Protein Expression level with Protein Sequence}
In this study, we embedded protein sequences using LLMs as a protein sequence feature extractor. ProtGPT2 \cite{ferruz2022protgpt2} is based on the GPT-2 \cite{Radford2019LanguageMA} architecture, which leverages the autoregressive transformer model to generate protein sequences and their latent representations specifically designed for protein sequences.

The extracted protein sequences in FASTA format are converted into tokenized inputs suitable for the ProtGPT2 model. Each protein sequence $S = \{s_1, s_2, \dots, s_L\}$, where $L$ is the length of the sequence and $s_i \in \mathcal{A}$ represents amino acids in the protein sequence, is tokenized into discrete elements using a specialized vocabulary derived from the amino acid alphabet $\mathcal{A}$ \cite{ferruz2022protgpt2}.

ProtGPT2 uses a multi-layer transformer architecture to embed these sequences. Each tokenized amino acid \( s_i \) is mapped to a learned embedding vector \( \mathbf{e}(s_i) \in \mathbb{R}^d \), where \( d \) is the dimensionality of the embedding space. The full sequence is represented as \cite{Attention, ferruz2022protgpt2}:
\[
\mathbf{E}(S) = \{\mathbf{e}(s_1), \mathbf{e}(s_2), \dots, \mathbf{e}(s_L)\}
\]

Since transformers are permutation-invariant, positional encodings $PE_i$ are added to each embedding vector to capture the sequential nature of the protein. The positional encoding used in transformer architectures involves sine and cosine functions that introduce unique positional values to each input embedding. This results in the final input embedding:
\[
\mathbf{z}_i = \mathbf{e}(s_i) + \mathbf{PE}_i
\]
ProtGPT2 then applies the self-attention mechanism to capture contextual relationships between amino acids within the protein sequence. For each attention head \( h \) the attention weights are computed as \cite{Attention, ferruz2022protgpt2}:
\[
\alpha_{ij}^h = \frac{\exp\left( \mathbf{q}_i^h \cdot \mathbf{k}_j^h / \sqrt{d_h} \right)}{\sum_{l=1}^{L} \exp\left( \mathbf{q}_i^h \cdot \mathbf{k}_l^h / \sqrt{d_h} \right)}
\]
where \( \mathbf{q}_i^h \) and \( \mathbf{k}_j^h \) are the query and key vectors for positions of amino acids \( i \) and \( j \), respectively, and \( d_h \) is the dimensionality of each attention head.

The attention output for each amino acid position \( i \) is computed as \cite{Attention, ferruz2022protgpt2}:
\[
\mathbf{z}_i' = \sum_{l=1}^{L} \alpha_{il}^h \mathbf{v}_l^h
\]
where \( \mathbf{v}_l^h \) is the value vector for amino acid position \( l \).

After the attention mechanism, a feedforward neural network (FFN) is applied, which consists of two linear transformations with a ReLU activation function in between. The final output embedding for each protein sequence is a combination of the attention mechanism’s output and the feedforward network, which results in a high-dimensional representation \( \mathbf{Y}(S) \in \mathbb{R}^{L \times d} \) capturing both local and global contextual information about the sequence \cite{Attention, ferruz2022protgpt2}. 

To get a sequence-level embedding, ProtGPT2 aggregates the individual token embeddings using the final hidden state of the first token. This results in a fixed-dimensional embedding vector \( \mathbf{D_S} \in \mathbb{R}^d \). The final hidden state of ProtGPT2 is \( d=1280 \) \cite{Radford2019LanguageMA, ferruz2022protgpt2}.

Each protein sequence $p_n$ (previously notated as S) is represented as a high-dimensional embedding vector $\mathbf{D_{p_n}} \in \mathbb{R}^{1280}$, where the vector captures functional, structural, and biochemical information about the protein. The embedding for protein $p_n$, where $n$ is the protein number (total of  10,625) can be expressed as:

\[
\mathbf{D_{p_n}} = [d_{p_{n_1}}, d_{p_{n_2}}, \dots, d_{p_{n_{1280}}}]
\]

where $d_{p_{n_f}}$ represents the embedding feature corresponding to the $f$-th dimension of protein \( n \).

The expression level for protein $p_n$ as \( \lambda_{p_n} \in \mathbb{R}^1 \), where $\lambda_{p_n}$ represents the protein's abundance measured through experimental techniques.

The integration of sequence embeddings $\mathbf{D_{p_n}}$ with the expression level $\lambda_{p_n}$ is achieved through element-wise multiplication. Mathematically, this can be expressed as:
\[
\mathbf{T_{p_n}} = \lambda_{p_n} \cdot \mathbf{D_{p_n}} = [\lambda_{p_n} \cdot d_{p_{n_1}}, \lambda_{p_n} \cdot d_{p_{n_2}}, \dots, \lambda_{p_n} \cdot d_{p_{n_{1280}}}]
\]
Each embedding feature $d_{p_{n_f}}$ is scaled by the expression level $\lambda_{p_n}$. This operation modulates the importance of each feature in the protein's embedding based on its biological abundance.

Finally, to address the high dimensionality of the protein data, where each protein is represented by an embedding of 1280 dimensions and each patient has data for 10,625 proteins, we applied Principal Component Analysis (PCA) to reduce the dimensionality of each protein embedding from 1280 to 1.

\[
\mathbf{F_{p_n}} = PCA(\mathbf{T_{p_n}})
\]

where $\mathbf{F_{p_n}} \in \mathbb{R}^1$ is the reduced representation of protein $p_n$, capturing the dominant variance from the scaled embedding vector.

\subsection{Clustering and Network Analysis}
We applied ensemble K-means clustering to group patients by integrating protein expression level and sequence with age, tumor size, and other clinical data. Ensemble clustering involves running the K-means clustering algorithm multiple times with different initialization settings to explore the stability of clusters and mitigate the effect of random initializations. The combination of proteomic data with clinical data creates a comprehensive representation for clustering. $X_m$ is the combination of the proteomic data with clinical data for the $m$-th patient. Then, it fed to the ensemble K-means clustering. The formula for the K-means objective function remains \cite{IKOTUN2023178}:
\[
\min \sum_{k=1}^{K} \sum_{\mathbf{X_m} \in CL_k} \|\mathbf{PA_m} - \boldsymbol{\mu_k}\|^2
\]

where $\boldsymbol{\mu_k}$ is the centroid of cluster $CL_k$, and $\mathbf{PA_m}$ represents the feature vector for the $m$-th patient. 

In the next step, we applied the consensus matrix to visualize the consistency across different clustering runs. It is a symmetric matrix where the entry $CM_{qv}$ indicates how often patients q and v were assigned to the same cluster across different clustering results. The consensus matrix is defined as:
\[
C_{qv} = \frac{1}{O} \sum_{o=1}^{O} \delta_{qv}^{(o)}
\]

where $O$ is the number of clustering solutions and $\delta_{ij}^{(o)} = 1$ if points $q$ and $v$ belong to the same cluster in the $o$-th solution, and 0 otherwise. This matrix provides a measure of how consistently pairs of patients are clustered together across all runs. 


\subsection{Classification}
Different classification algorithms such as XGBoost and Random Forest are applied to classify the patient data to predict different aspects like survival of patients. 
One of these predictions is to predict the status of three biological markers, ER, PR status, and HER2 Final Status, which are essential in breast cancer classification. Each marker is a target variable, a multi-output classification problem. The $X_m$ is used to predict them as a multi-label. The Hamming loss metric measures the fraction of incorrectly predicted labels, providing an overall view of the model’s ability to classify each target accurately. Lower Hamming loss indicates better classification performance. The formula for Hamming loss is given by \cite{NEURIPS2020_20479c78}:
\[
HL = \frac{1}{m \times b} \sum_{q=1}^{m} \sum_{u=1}^{b} 1(y_{qu} \neq \hat{y}_{qu})
\]

Where $m$ is the number of patients (77), $b$ is the number of labels per patient. $y_{qu}$ is the actual value of the $u$-th label for the $q$-th patient. $\hat{y}_{qu}$ is the predicted value of the $u$-th label for the $q$-th patient. $1(\cdot)$ is the indicator function, which is 1 if the condition inside is true and 0 otherwise.

\section{Results}
\subsection{Classifications}
To evaluate the effectiveness and robustness of integrating protein sequence data with protein expression levels, we applied different machine learning algorithms to predict the survival outcomes of breast cancer patients. The survival status, defined as either "alive" or "deceased", was predicted based on a comprehensive feature set comprising the integrated protein sequence and expression level and relevant clinical data such as tumor size and patient age. The classification model demonstrated a strong performance on the XGBoost, achieving an F1 score of 0.88. This result underscores the model's ability to distinguish between patients who are likely to survive and those at a higher risk.

Then, We employed multiple classification algorithms to evaluate the predictive capability of our integrated protein sequence and expression level in determining the status of breast cancer biomarkers, including ER, PR, and HER2 status. We treated the prediction of these three markers as a multi-output classification problem, where each marker represents an individual target to be predicted simultaneously. Table \ref{tab:classification_metrics_markers} compares various classification algorithms, showcasing the precision, recall, F1 score, and Hamming loss for each model. These metrics are reported as weighted averages, reflecting the model's ability to predict multiple target labels across all patient samples effectively.

\begin{table}[h]
    \centering
    \begin{tabular}{lcccc}
        \hline
        \textbf{Algorithm} & \textbf{Precision} & \textbf{Recall} & \textbf{F1 score} & \textbf{Hamming Loss} \\ 
        \hline
         K-Nearest Neighbors & 0.64 & 0.79 & 0.69 &  0.31 \\
         Support Vector Machine & 0.67 & 0.83 & 0.74 & 0.19 \\
         Kernel SVM &  0.67 & 0.83 & 0.74 & 0.19 \\
        Random Forest  & 0.72 & 0.79 & 0.75 & 0.17 \\
        Gradient Boosting  & 0.72 & 0.79 & 0.75 &  0.17 \\
        AdaBoost & 0.88 & 0.79 & 0.81 & 0.17 \\
        CatBoost & 0.89 & 0.83 & 0.82 & 0.15 \\
        XGBoost & 0.88 & 0.88 & 0.87 & 0.18\\
        \hline
    \end{tabular}
    \caption{Precision, recall, and F1 score for predicting all markers as a multi-output classification problem based on different algorithms reported in the weighted average}
    \label{tab:classification_metrics_markers}
\end{table}

The results in Table \ref{tab:classification_metrics_markers} show that XGBoost achieved the best overall performance, with an F1 score of 0.87 and balanced precision and recall values of 0.88 each. This indicates that XGBoost was highly influential in predicting the multi-output targets accurately, achieving strong generalization across all three biomarkers while maintaining a relatively low Hamming loss of 0.18. CatBoost and AdaBoost also performed well, with CatBoost showing a slight edge in precision and Hamming loss compared to XGBoost.

To further investigate which proteins were most influential in predicting the biomarker statuses, we analyzed the feature importance scores from the XGBoost model. The integration of protein expression levels and sequence information was used for each protein, allowing the model to leverage both the abundance of the proteins and their functional sequence-derived attributes. Table \ref{tab:feature_importance} lists the top 10 proteins, including critical proteins with the highest importance scores during classification. Among the proteins, KMT2C and GCN1 were the most influential proteins, with importance scores of 0.119 and 0.114, respectively. 

\begin{table}[h]
    \centering
    \begin{tabular}{lcccccccccc}
        \hline
        \textbf{Proteins} & KMT2C & GCN1 & CLASP2 & AQR & MYO1B & MAP2 & TTN & CLIP1 & SYNE2 & TPM1 \\
        \textbf{Importance Score} & 0.119 & 0.114 & 0.080 & 0.075 & 0.074 & 0.062 & 0.056 & 0.053 & 0.049 & 0.038 \\
        \hline
    \end{tabular}
    \caption{Importance score values for top 10 important proteins from XGBoost model}
    \label{tab:feature_importance}
\end{table}

\begin{figure}
  \centering
  \includegraphics[width=0.7\linewidth]{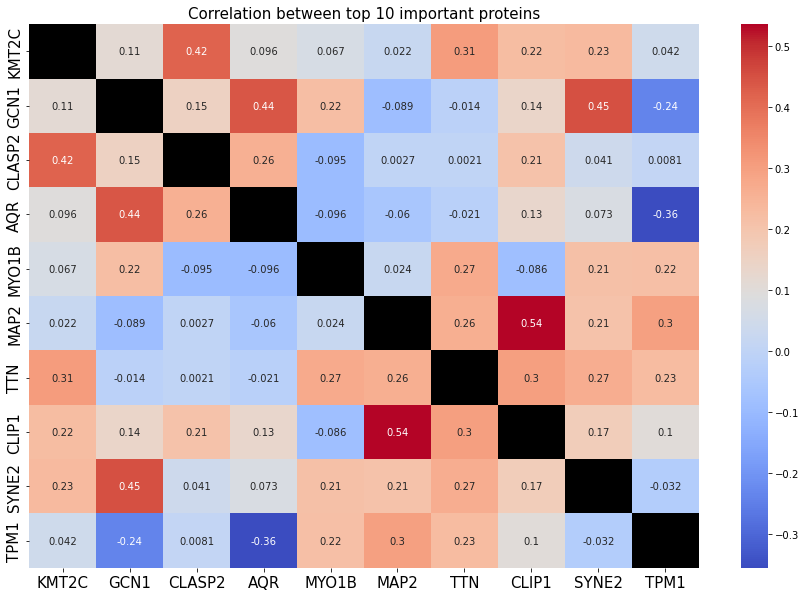}
  \caption{Pearson correlation matrix of the top 10 important proteins in the biomarkers prediction}
  \label{fig:Correlations}
\end{figure}

Figure \ref{fig:Correlations} illustrates the correlation matrix for the essential proteins from the XGBoost results. This matrix illustrates the pairwise Pearson correlation coefficients among the top proteins, providing insight into how these proteins relate. The matrix reveals several patterns of interdependence between specific proteins, with varying degrees of positive and negative correlations. The intensity of the color represents the strength and direction of the correlation; red tones signify strong positive correlations, while blue tones indicate strong negative correlations. The interrelationships among proteins help identify groups of proteins that could be contributing jointly to disease pathology or acting through similar pathways. For instance proteins like MAP2 and CLIP1, AQR and TPM1, the high correlation between some proteins (either negative or positive) suggests they may co-regulate or participate in similar cellular processes, which could be biologically significant in distinguishing different breast cancer subtypes or outcomes.

\begin{figure}
  \centering
  \includegraphics[width=0.7\linewidth]{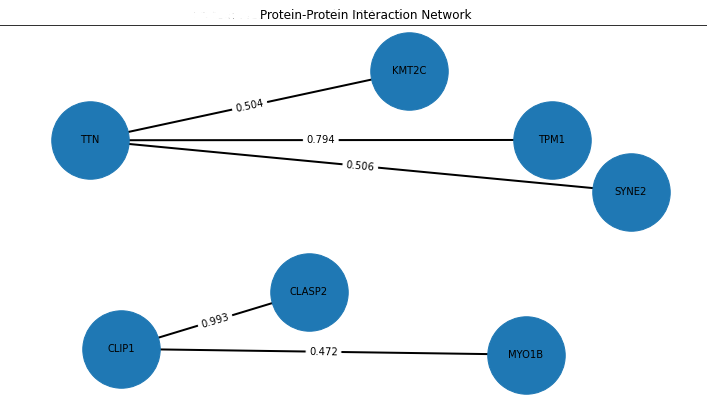}
  \caption{Protein-protein interactions based on the STRING-db of the top 10 important proteins}
  \label{fig:PPI}
\end{figure}

Figure \ref{fig:PPI} presents the protein-protein interaction (PPI) network for the top 10 most important proteins identified from the XGBoost feature importance analysis. This network was constructed using STRING-db \cite{string_db2023}, a well-known database for retrieving known and predicted protein-protein interactions. The nodes in the network represent individual proteins, and the edges (lines connecting the nodes) indicate interactions between these proteins. The values along the edges denote the interaction confidence scores, which range from 0 to 1, reflecting the strength or likelihood of an interaction between each pair of proteins. From the network, we observe several high-confidence interactions. CLIP1 and CLASP2 exhibit an extremely high interaction score of 0.993, suggesting a robust functional relationship. This could indicate that these two proteins are part of a shared pathway or play complementary roles in cellular processes.
TTN interacts with multiple proteins, such as KMT2C (interaction score 0.504), TPM1 (score 0.794), and SYNE2 (score 0.506). This suggests that TTN may have a central role in connecting multiple proteins, indicating a potential influence on several pathways in cancer biology.
Another notable interaction is between CLIP1 and MYO1B, with an interaction score of 0.472, which suggests a moderate but biologically significant relationship.
The PPI network provides insights into the complex biological interactions and the functional relationships between these proteins, potentially revealing collaborative roles in breast cancer pathogenesis. 

\subsection{Clustering}

\begin{figure}
  \centering
  \includegraphics[width=0.8\linewidth]{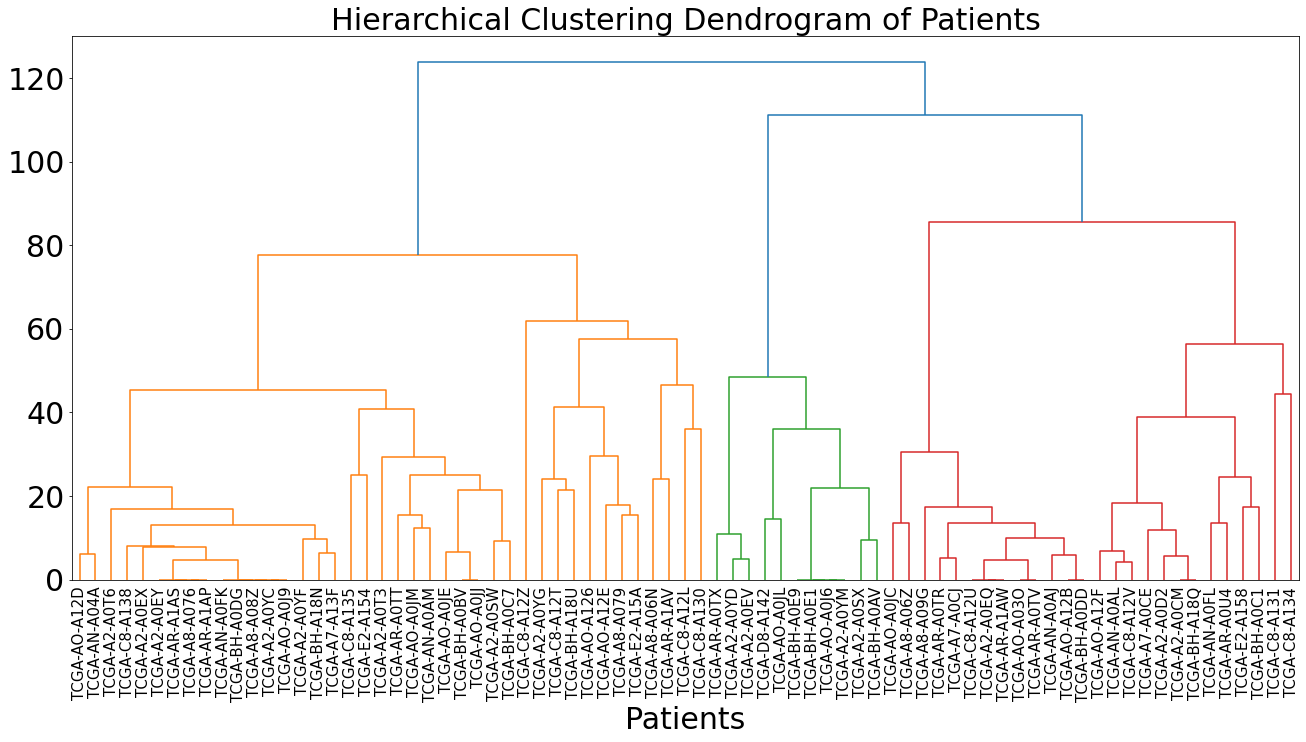}
  \caption{Dendrogram resulting from the ensemble clustering of the patient dataset combined with hierarchical clustering for each patient case}
  \label{fig:dendrogram}
\end{figure}

Figure \ref{fig:dendrogram} presents the dendrogram resulting from ensemble clustering of the patient dataset, with hierarchical clustering applied for visualization. The dendrogram reveals a clear hierarchical structure, separating patients into three major color-coded clusters: orange, green, and red. These clusters indicate well-defined patient subgroups based on consensus similarity patterns, with nested sub-branches reflecting additional fine-scale subgroupings. Shorter linkage distances between branches represent higher similarity, while the distinct separation between color groups underscores strong intergroup differences.

To further explore the biological relevance of these clusters, we analyzed the co-occurrence of three key breast cancer markers of ER, PR, and HER2 within each group, as shown in Appendix Figure \ref{fig:cooc_matrices}. In the orange cluster, ER and PR status frequently occur together, indicating a moderate correlation between these markers. In contrast, ER status and HER2 status and PR status and HER2 status tend to appear in opposite states, suggesting a negative relationship between these markers. In the green cluster, ER and PR status are perfectly correlated, meaning they are always present in the same state (True or False) within this group. ER and HER2 status, along with PR and HER2 status, also tend to occur together, but this correlation is less consistent compared to ER and PR. Finally, the red cluster shows ER and PR status exhibit a strong correlation, often appearing in the same state. However, ER and HER2 status and PR and HER2 status show minimal to no correlation, suggesting their occurrences are mainly independent of each other within this group.

\section{Discussion}

ER and PR are both hormone receptors, and their presence typically characterizes hormone receptor-positive (HR+) breast cancers. ER is involved in regulating the expression of the PR gene. When estrogen binds to its receptor (ER), this often leads to upregulating PR expression. Therefore, when breast cancer is ER-positive, it is very likely also to be PR-positive \cite{li2022role}. This results in a high co-occurrence or perfect correlation between these two markers in many clusters, as seen in the second and third clusters.
ER-positive breast cancers are generally dependent on estrogen signaling for growth and survival, and they often express PR as a downstream effect. The presence of both ER and PR usually indicates a tumor that responds well to hormone therapy, such as tamoxifen or aromatase inhibitors \cite{li2022role}.

The relationships between ER/PR and HER2 are more complex, often showing a tendency for negative correlation. ER-positive and HER2-positive statuses are often mutually exclusive because they represent different molecular subtypes of breast cancer. ER/PR-positive tumors are usually tend to be more hormone-dependent and less aggressive compared to HER2-positive subtypes \cite{onitilo2009breast}.
HER2-positive breast cancers tend to grow independently of hormone signaling. Instead, they rely on overexpression of the HER2 protein, which promotes cell growth via the MAPK and PI3K/AKT pathways \cite{cheng2024comprehensive}. This reliance on a different growth mechanism often leads to a negative relationship between HER2 and hormone receptors \cite{onitilo2009breast}. In the first cluster, ER and HER2, as well as PR and HER2, often appear in opposite states.
As shown in Appendix Figure \ref{fig:cooc_matrices}, In the first cluster, ER and PR are positively correlated, while both show moderate negative correlations with HER2 expression. This inverse relationship between hormone receptors and HER2 is consistent with immunohistochemistry (IHC) based breast cancer subtypes, where ER/PR+ tumors (Luminal A and B) are generally distinct from HER2-enriched or triple-negative tumors \cite{onitilo2009breast}. In the second cluster, the perfect correlation between ER and PR could indicate a cluster primarily composed of luminal A breast cancers, which typically show both strong ER and PR expression. The variable relationship with HER2 may reflect a small subset of luminal B cancers or suggest that HER2-positive tumors in this cluster co-express ER/PR, although not consistently. The last cluster suggests ER and PR are generally linked, but HER2 is mainly independent. It could represent a group with almost exclusively hormone receptor-positive, HER2-negative tumors. These tumors may be purely luminal A subtype, where hormone receptor positivity is dominant, and HER2 expression is largely absent. The analysis reveals an overall inverse relationship between hormone receptors (ER/PR) and HER2, consistent with known breast cancer subtypes like Luminal A and HER2-enriched. Notably, the second cluster shows perfect ER/PR correlation with variable HER2 expression, which may represent a hybrid or unusual luminal B subgroup. This pattern suggests a potential novel cluster characterized by co-expression of hormone receptors and moderate HER2, warranting further investigation into its molecular features and clinical behavior.

To check the proteins represented in the correlation matrix in Figure \ref{fig:Correlations} and their potential connections to the ER, PR, and HER2 status in breast cancer, it is essential to explore their involvement in cellular mechanisms relevant to breast cancer's onset, progression, and differentiation. KMT2C is involved in chromatin remodeling, and has been shown to promote estrogen receptor (ER$\alpha$) function in ER-positive, HER2-negative breast cancer. Mutations in KMT2C may disrupt hormone receptor signaling and are associated with aggressive luminal subtypes such as luminal A and B \cite{fagan2019compass}. GCN1 regulates protein synthesis during amino acid starvation by facilitating GCN2 activation, which in turn phosphorylates eIF2$\alpha$ to reduce global translation. While its direct influence on hormone receptor status is not reported, its modulation of stress response pathways may indirectly impact hormone receptor expression and survival in hormone-dependent cancers such as breast cancer \cite{sattlegger2005polyribosome}. CLASP2, a microtubule plus-end tracking protein, regulates microtubule dynamics essential for cell migration. Although direct evidence in HER2-positive cancers is limited, its overexpression in other malignancies has been shown to enhance focal adhesion turnover and epithelial-to-mesenchymal transition (EMT), suggesting a potential role in promoting invasive behavior and metastatic progression in HER2-driven tumors. \cite{wattanathamsan2022emerging_clasp2}. AQR is a spliceosomal factor essential for TNBC cell proliferation. By regulating splicing, it may indirectly influence hormone receptor expression and contribute to breast cancer heterogeneity \cite{koedoot2021splicing}. MYO1B, a motor protein regulated through SRSF1-mediated alternative splicing, contributes to drug resistance, proliferation, and invasion in breast cancer cells, including hormone receptor-positive and triple-negative subtypes \cite{my01b_2025}. MAP2 stabilizes microtubules  \cite{dehmelt2005map2}, may potentially affecting cell response to hormone receptor or HER2 signaling. SYNE2 links the cytoskeleton to the nucleus via the LINC complex and regulates p21 expression independently of TP53, suggesting a role in structural stability and cell cycle control in luminal breast cancers \cite{han2016egfr}. TPM1 drive EMT transition and chemoresistance via actin remodeling in ovarian cancer and may similarly influence hormone receptor localization and subtype distinction in breast cancer \cite{xu2024tropomyosin1}. Therefore, several proteins are linked to hormone receptor signaling, cytoskeletal remodeling, and EMD transition, suggesting roles in subtype differentiation. Their potential influence on ER, PR, and HER2 expression warrants further clinical investigation to clarify their relevance in breast cancer progression and heterogeneity.


PPIs in breast cancer often indicate shared biological pathways that regulate critical cellular functions, including transcription regulation, cytoskeletal dynamics, and cell motility, affecting the expression and function of ER, PR, and HER2. Specifically, PPIs among proteins in the dataset point towards a collective role in coordinating receptor signaling and influencing breast cancer subtypes \cite{kim2021protein}. The interaction network reveals two distinct functional modules. The first, centered on TTN, connects chromatin remodeling (via KMT2C) with cytoskeletal stability (TPM1 and SYNE2), suggesting a nuclear structural axis that may regulate hormone receptor function and support luminal subtype differentiation \cite{fagan2019compass, xu2024tropomyosin1, han2016egfr}. The second module, linking CLIP1, CLASP2, and MYO1B, highlights a microtubule motor axis associated with migration, EMT, and potential therapy resistance \cite{wattanathamsan2022emerging_clasp2, my01b_2025}, may particularly in HER2-positive and triple-negative breast cancers. The strong interaction between CLIP1 and CLASP2 (interaction score: 0.993) suggests tightly coordinated roles in cytoskeletal remodeling and structural maintenance, which are critical for tumor progression and metastasis \cite{wattanathamsan2022emerging_clasp2}. This network highlights how individual proteins, prioritized by the XGBoost model, may functionally converge or influence each other within the broader landscape of tumor biology. Understanding these interactions provides valuable insights into the mechanisms driving tumor progression and may inform the development of targeted therapeutic strategies that account for these molecular relationships.

Integrating large language models like ProtGPT2 into bioinformatics allows researchers to capture nuanced protein sequence features, combining these with expression levels to create contextually relevant embeddings. In bioinformatics, protein expression levels are critical for determining a protein's biological impact in a cell or tissue. Mathematically, this involves multiplying the embedding vector, E, of a protein sequence generated by ProtGPT2 by the corresponding protein expression level ($\lambda_{p_n}$), resulting in a new scaled embedding. This operation effectively scales each element of the embedding vector based on the protein's expression level, introducing biological significance directly into the numerical representation. This scaling can be seen as a feature-specific weight that adjusts the impact of the embedding in downstream analyses, such as clustering or classification models, ensuring biologically essential proteins are given appropriate emphasis in machine learning models.
Gating mechanisms are used in neural networks to modulate information processing, as seen in LSTMs or attention-based models like transformers  \cite{LSTM, Attention}. Multiplying the protein embedding vector by the expression level acts like a gating mechanism, where the expression level $\lambda_{p_n}$ serves as a gate that either amplifies or reduces the influence of the embedding based on the protein's biological importance. When the expression level is high, more information passes through, akin to a gate being "opened" wider for influential inputs. In contrast, a lower expression level reduces the impact of that protein. Using this biologically-informed embedding strategy, our method produced promising results in both clustering and classification tasks. Notably, it enabled the identification of biologically relevant protein clusters and revealed potential subtype-specific protein signatures, demonstrating its utility in uncovering meaningful patterns within complex omics data.

\section{Conclusion}
This study introduces a novel framework for breast cancer analysis by integrating protein sequence with expression data using ProtGPT2-derived embeddings, and enhanced with machine learning models such as XGBoost for subtype classification and predict clinical outcomes. By combining structural protein information with quantitative expression levels, this study creates a comprehensive representation of the proteins involved, improving accuracy in predicting key biomarkers (ER, PR, HER2) and enabling the clustering of patients into meaningful subgroups. Ensemble clustering revealed the molecular heterogeneity of breast cancer, uncovering patterns aligned with known subtypes and suggesting avenues for targeted therapies. Additionally, protein–protein interaction analysis reveals meaningful functional relationships, such as those involving proteins like KMT2C, linked to estrogen receptor regulation, and MYO1B, which contributes to drug resistance, and invasion in hormone receptor-positive and triple-negative breast cancers, highlighting their potential as therapeutic targets within a biologically informed framework. Overall, this integrated approach demonstrates the power of combining advanced protein language models with expression-informed embeddings, enhanced by machine learning, to advance precision oncology and provide a framework that can be extended to other complex diseases for deeper biological insights and more effective personalized treatments.



\bibliographystyle{unsrt}  
\bibliography{references}  

\appendix
\renewcommand{\thefigure}{A.\arabic{figure}} 
\setcounter{figure}{0} 

\section{Supplementary Figures}

\begin{figure}[htbp]
    \centering

    \subfigure[Co-occurrence matrix of the Orange Cluster]{
        \includegraphics[width=0.7\textwidth]{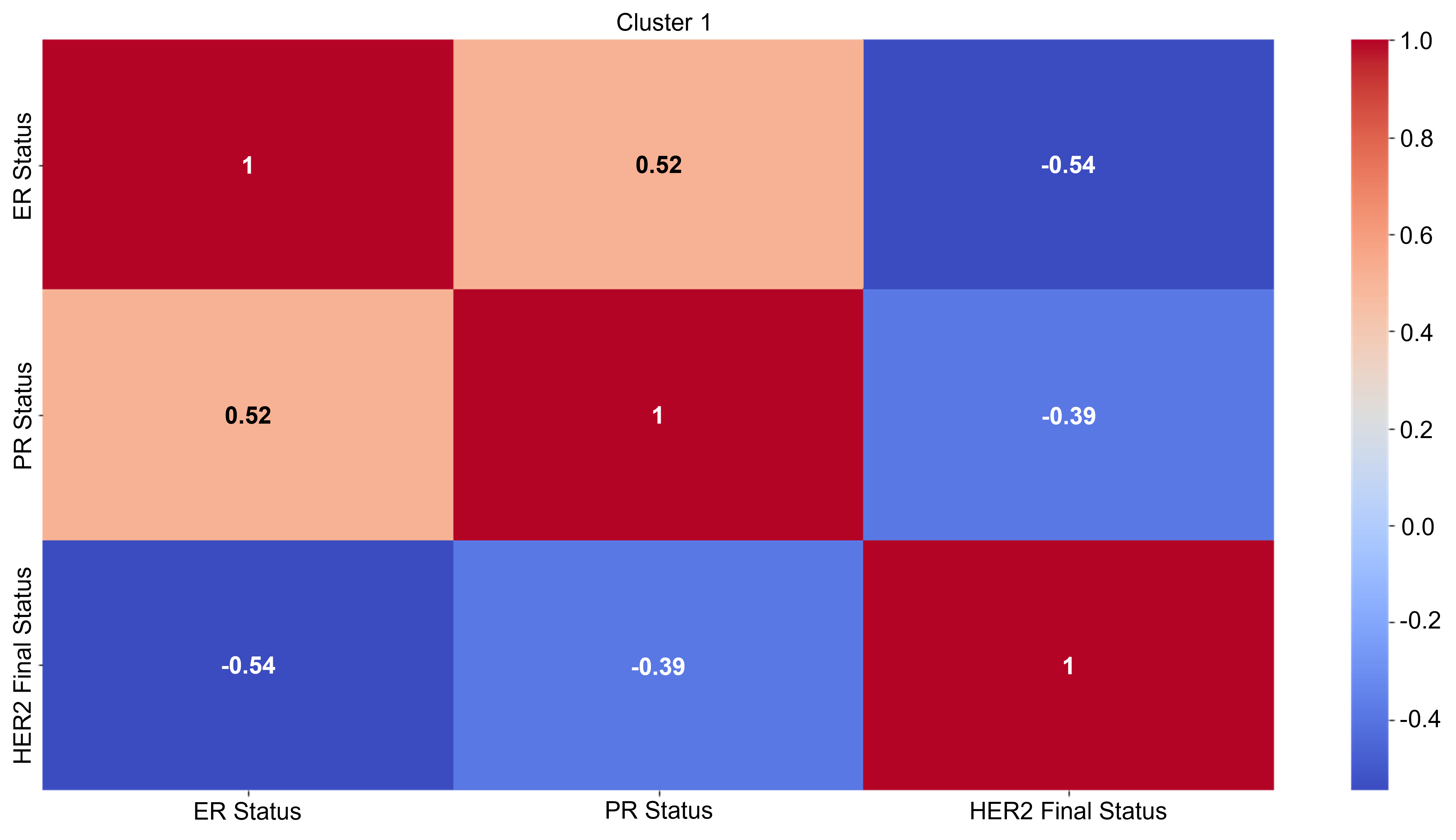}
        \label{fig:cluster1}
    }

    \subfigure[Co-occurrence matrix of the Green Cluster]{
        \includegraphics[width=0.7\textwidth]{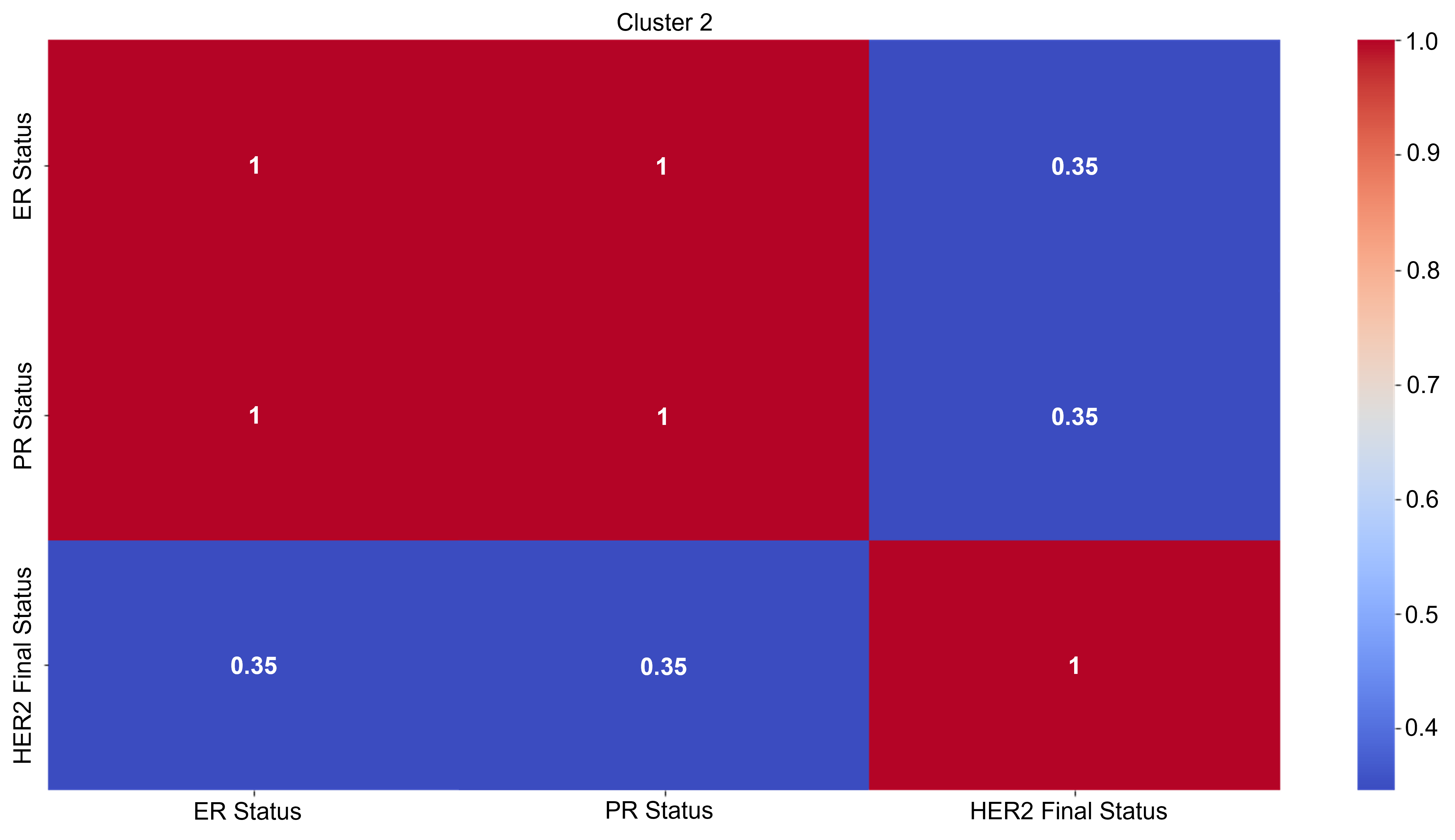}
        \label{fig:cluster2}
    }

    \subfigure[Co-occurrence matrix of the Red Cluster]{
        \includegraphics[width=0.7\textwidth]{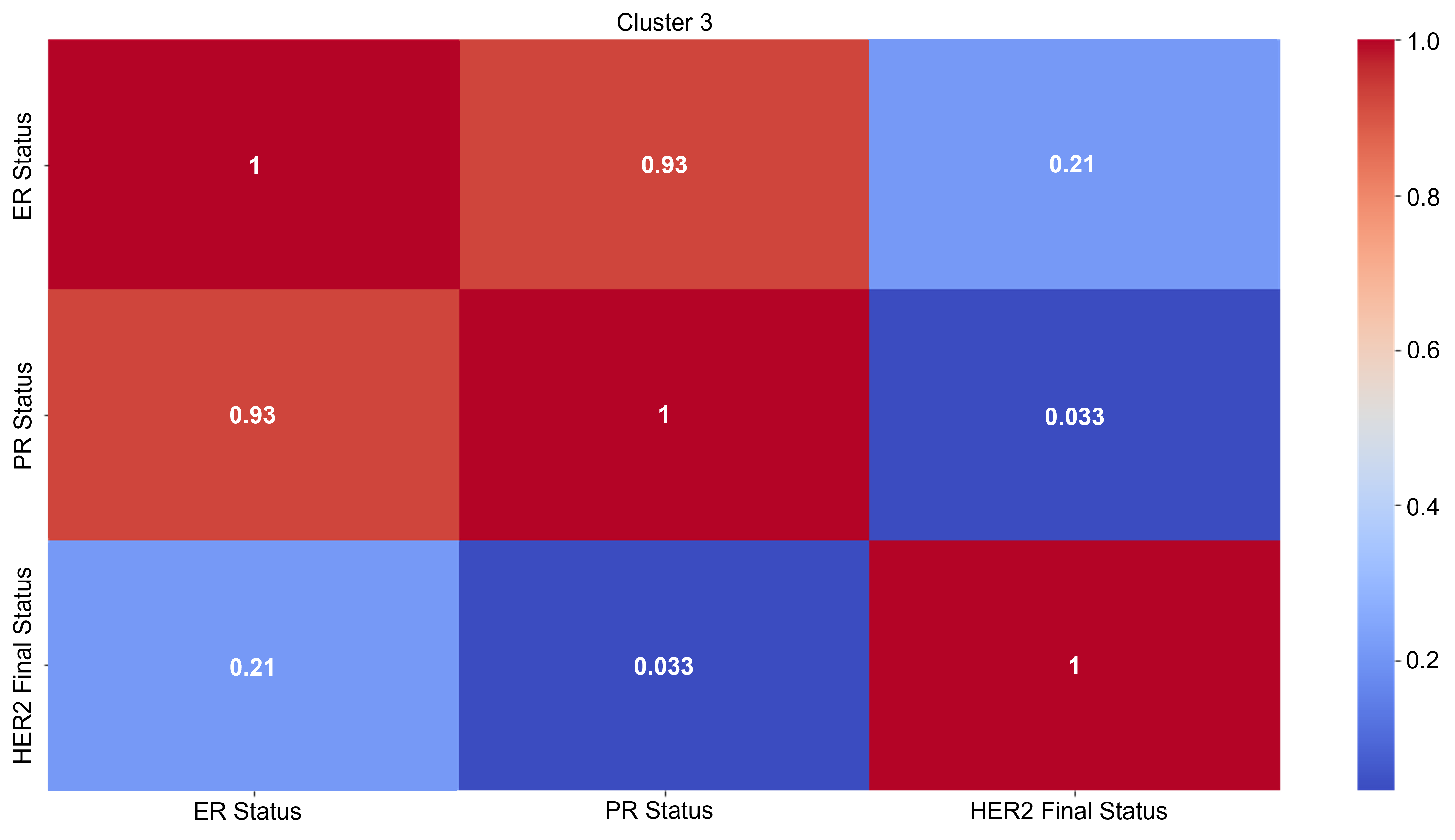}
        \label{fig:cluster3}
    }

    \caption{Co-occurrence matrices for three patient clusters (Orange, Green, and Red), illustrating correlations between breast cancer markers.}
    \label{fig:cooc_matrices}
\end{figure}

\end{document}